\documentclass[conference]{IEEEtran}
\IEEEoverridecommandlockouts
\usepackage{verbatim}
\usepackage{longtable}
\usepackage{adjustbox}
\usepackage{listings}
\usepackage{xcolor}

\lstdefinelanguage{Solidity}{
  morekeywords={pragma, solidity, contract, struct, event, function, mapping, address, bytes, bytes32, bool, external, public, require, emit, calldata, payable, return, returns, if, else, memory, keccak256},
  sensitive=true,
  morecomment=[l]{//},
  morecomment=[s]{/*}{*/},
  morestring=[b]"
}

\usepackage{amsmath}
\usepackage{amsfonts}
\usepackage{amssymb}
\usepackage{multirow}
\usepackage{amsmath,amssymb,amsfonts}
\usepackage{fancyhdr}

\usepackage{graphicx}
\usepackage{textcomp}
\usepackage{xcolor}
\usepackage{enumitem}
\usepackage{float}
\usepackage{graphicx}
\usepackage[normalem]{ulem}
\usepackage{url}
\useunder{\uline}{\ul}{}
\usepackage{cite}
\usepackage{amsmath,amssymb,amsfonts}
\usepackage{graphicx}
\usepackage{textcomp}
\usepackage{comment}
\usepackage{subcaption}

\usepackage{xcolor}
\def\BibTeX{{\rm B\kern-.05em{\sc i\kern-.025em b}\kern-.08em
    T\kern-.1667em\lower.7ex\hbox{E}\kern-.125emX}}
\usepackage{algorithm}
\usepackage[noend]{algpseudocode}

\begin{document}

\title{Towards Secure and Trustworthy DAOs for Cross-Chain Governance}

\author{\IEEEauthorblockN{Faisal Haque Bappy, Tahrim Hossain, Tarannum Shaila Zaman, Tariqul Islam}
\IEEEauthorblockA{
University of Maryland Baltimore County\\
Email: \{fbappy1, m482, zamant, mtislam\}@umbc.edu} 
}
\maketitle

\begin{abstract}
Cross-chain DAOs face unique security challenges that go beyond traditional single-chain vulnerabilities. This paper identifies and categorizes four critical attack vectors in cross-chain DAO governance: bribery attacks, token control exploits, human-computer interaction deceptions, and protocol vulnerabilities. We propose a comprehensive security framework with a multi-layered architecture that integrates cryptographic trust anchors, fraud-resistant consensus mechanisms, and decentralized validation techniques to address these threats. Our framework introduces novel components, including a Governance Kernel with on-chain rule verification, a Cross-Chain Trust Layer using threshold cryptography, and a Resilience Layer offering time-locked decision reversals and progressive dispute resolution. By establishing a structured set of countermeasures, this work lays the foundation for secure, transparent, and attack-resistant governance across diverse blockchain environments.
\end{abstract}


\section{Introduction}
Decentralized Autonomous Organizations (DAOs) play a key role in blockchain governance, enabling transparent, code-enforced collective decision-making. In homogeneous blockchain environments, DAOs have proven effective in community-driven project management\cite{hassan2021decentralized}, treasury allocation\cite{schneider2022decentralized}, protocol upgrades\cite{van2024upgradeable}, and dispute resolution without centralized intermediaries\cite{guillaume2023blockchain}. This has driven their adoption in DeFi protocols, NFT communities, and blockchain infrastructure projects.

As blockchain ecosystems evolve, cross-chain communication methods are becoming increasingly important. These methods enable complex interactions between once separate networks, extending beyond basic token exchanges\cite{ou2022overview}. Seamless cross-chain governance is the next frontier, potentially allowing DAOs to coordinate resources and make decisions across multiple blockchain environments simultaneously. However, current DAO structures face significant security and trust vulnerabilities. Recent incidents\cite{belenkov2025sok, lee2023sok} have revealed weaknesses in governance mechanisms, including flash loan attacks\cite{qin2021attacking}, governance token manipulation\cite{feichtinger2024sok}, proposal bribery\cite{karakostas2024blockchain}, and smart contract vulnerabilities\cite{liao2024smartaxe}. These exploits have resulted in substantial financial losses and diminished confidence in DAO governance models. Moreover, centralization tendencies, despite theoretical decentralization, exacerbate these challenges, with voting power often concentrated among a small subset of participants.

In cross-chain environments, these issues are amplified by added complexity. Asynchronous communication introduces timing-related attack vectors\cite{augusto2024sok}, while inconsistent security models across chains and reliance on cross-chain bridges and oracles create new points of failure\cite{haugum2022security}. The lack of standardized governance protocols further leads to coordination inefficiencies, requiring new defensive approaches beyond single-chain security practices.

In this paper, we analyze the cross-chain attack vectors for DAOs and propose a theoretical framework for secure cross-chain DAO governance that addresses these emerging challenges through a structured set of countermeasures. Our framework introduces a multi-layered security model that integrates cryptographic trust anchors, fraud-resilient consensus mechanisms, decentralized notaries, and cross-chain validation techniques to enhance governance integrity. By leveraging threshold cryptography, multi-party computation (MPC), and decentralized auditing mechanisms, we provide a governance model that ensures security, transparency, and resistance to malicious manipulation. The following are the key contributions of this paper.

\begin{itemize}
    \item We systematically identify and categorize cross-chain governance attack vectors, including bribery, token control exploits, phishing, and smart contract vulnerabilities.
    \item We propose a novel governance framework incorporating cryptographic security, fraud-resistant consensus mechanisms, and dispute resolution strategies to mitigate emerging threats in cross-chain DAOs.
    \item We design a resilience layer that introduces time-locked decision reversals, MPC-based key custody, and multi-tier trust anchors to enhance security and trust in decentralized governance.
\end{itemize}

The rest of the paper is organized as follows: Section \ref{sec:background} provides the background and rationale for cross-chain DAOs, Section \ref{sec:attacks} explores the various attack vectors associated with cross-chain governance, Section \ref{sec:framework} presents a comprehensive governance framework for securing cross-chain DAOs, and Section \ref{sec:conc} concludes with a discussion on future research directions.

\section{Background and Rationale}
\label{sec:background}
DAOs are blockchain-based organizations that enable collective governance without hierarchical structures, distributing decision-making among stakeholders to mitigate centralization and enhance transparency \cite{liu2021technology}. Central to DAOs are smart contracts \cite{szabo1996smart}, which are self-executing codes that securely enforce decisions through immutable rules, ensuring decentralized governance controlled by members, not central authorities. Governance rights are typically represented by tokens, empowering holders to participate in decision-making \cite{calcaterra2023reputation}. Voting can take place either directly on-chain or off-chain, depending on the governance model \cite{monteiro2024exploring}. To address power imbalances inherent in simple majority voting, DAOs have moved beyond the basic ``one token, one vote" approach \cite{han2025review}, incorporating more nuanced mechanisms such as quadratic voting \cite{benhaim2024balancing}, conviction voting \cite{ding2023voting}, and tenure-based systems \cite{curve2020dao}. These innovations help ensure fairer representation and mitigate the risk of governance manipulation by large token holders. The flexibility and transparency of DAOs make them valuable across various domains, including investment, philanthropy, protocol governance, and digital asset management \cite{tang2025decentralised}. 

Despite their potential, DAOs face significant security and governance risks. Smart contract vulnerabilities and governance attacks have led to major exploits, such as The DAO hack \cite{dhillon2017dao} and the Beanstalk Protocol exploit \cite{feichtinger2024sok}. Recent studies on DAOs have examined the distribution of voting power \cite{fritsch2024analyzing}, its impact on governance decisions \cite{feichtinger2023hidden}, and the broader effects of centralization on DAO operations \cite{sharma2024unpacking}. A key governance challenge that has emerged is hidden vote buying, where governance power can be secretly traded through smart contracts, enabling Dark DAOs to manipulate decision-making processes while remaining undetected \cite{austgen2023dao}. This issue ties into the broader concept of Governance Extractable Value (GEV), which, similar to Miner Extractable Value (MEV), highlights how governance mechanisms can be exploited for financial gain \cite{ournetwork2025}. These findings underscore the risks of governance manipulation and exploitation in DAOs. 

As blockchain networks diversify, cross-chain DAOs such as Curve \cite{curve2020dao} and Moonbeam \cite{moonbeam} have emerged, enhancing governance interoperability across multiple chains. These frameworks, however, introduce complexities, security risks, and coordination challenges. Ensuring cross-chain compatibility requires substantial technical resources, as vulnerabilities in one chain can affect others. Governance models must evolve to manage diverse networks with robust mechanisms that extend beyond single-chain contexts. Furthermore, issues prevalent in single-chain DAOs, such as governance manipulation and voting imbalances, can manifest in new forms in cross-chain environments, necessitating innovative solutions to maintain security and decentralization.

\section{Cross-Chain Attack Vectors for DAO}
\label{sec:attacks}
\subsection{Bribing Attacks}
In homogeneous DAOs, bribing attacks occur through vote-buying, where attackers use financial incentives to sway governance. Flash loans worsen this by enabling temporary token acquisition to push malicious proposals \cite{qin2021attacking}. Platforms like Curve\cite{curve2020dao} has faced governance takeovers through such token manipulation. In cross-chain DAOs, bribery risks are heightened by fragmented governance and delayed decision execution. Attackers can exploit discrepancies across chains, temporarily acquiring tokens to manipulate decisions before security measures take effect. The lack of unified oversight enables attackers to accumulate influence without triggering alarms, making bribery a more potent threat in multi-chain environments.  

\begin{figure}[h]
  \centering
  \includegraphics[width=0.9\columnwidth]{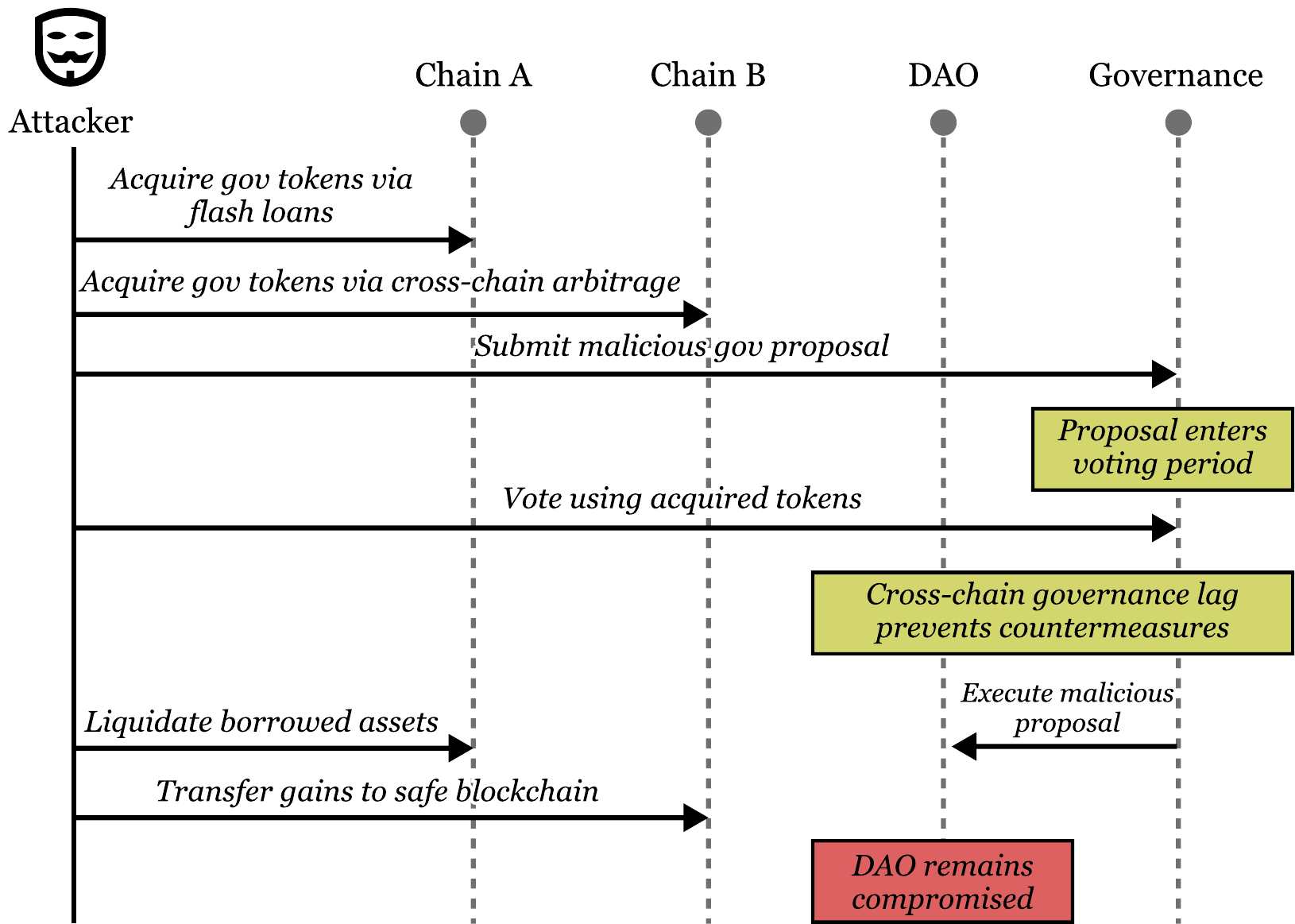}
  \caption{Cross-Chain Bribing Attack}
  \label{fig:bribing}
\end{figure}

A typical attack workflow (Figure \ref{fig:bribing}) begins with an attacker acquiring governance tokens through flash loans or cross-chain arbitrage. They then submit or support a governance proposal that benefits them, leveraging cross-chain governance lags to ensure that countermeasures cannot be implemented in time. Once the proposal passes and executes, the attacker exits by liquidating the borrowed assets or transferring gains to a safe blockchain, leaving the DAO compromised. The delays inherent in cross-chain consensus prevent rapid response, allowing the attack to succeed before defenses activate.  

\begin{figure}[!b]
  \centering
  \includegraphics[width=0.9\columnwidth]{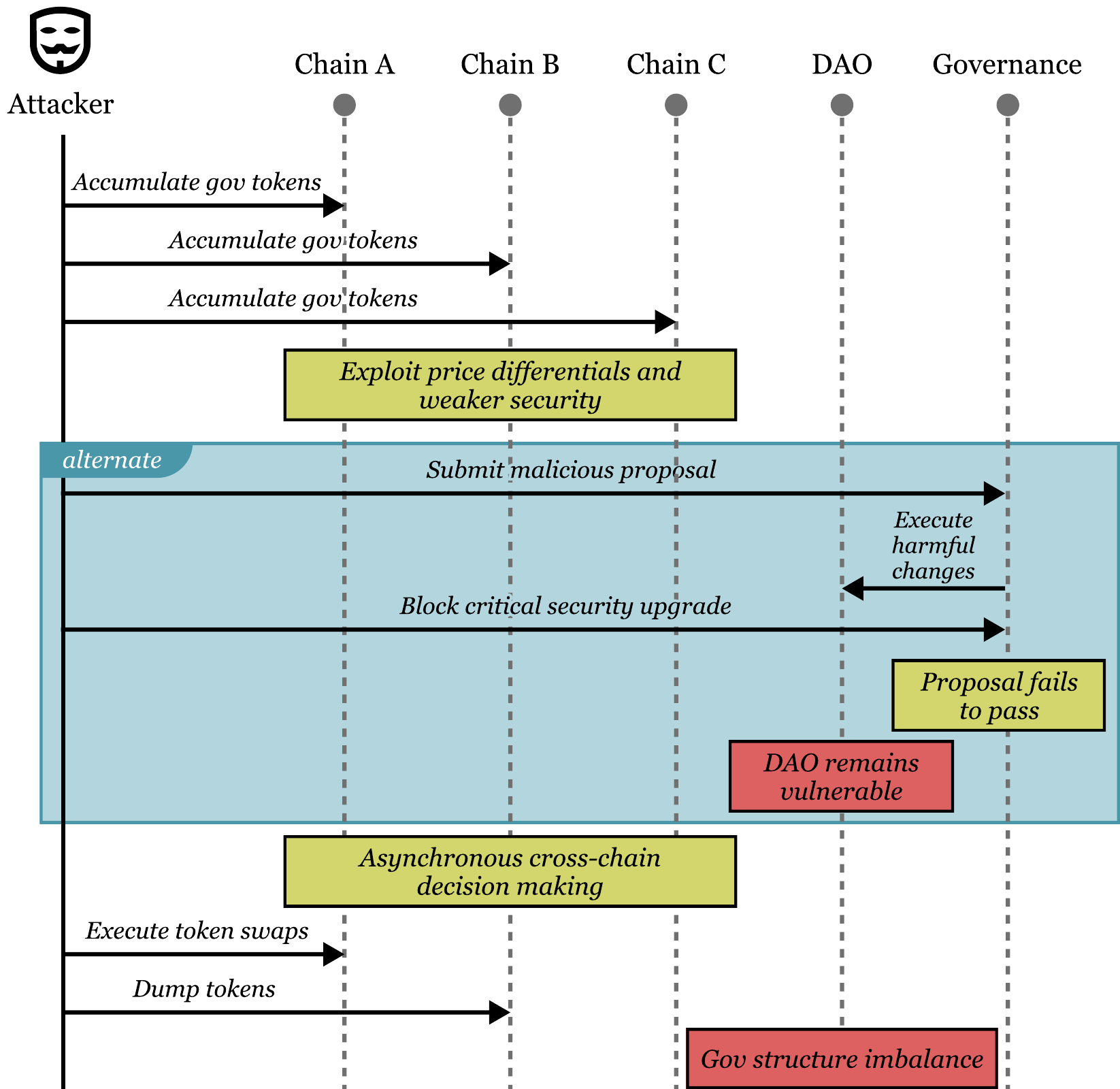}
  \caption{Cross-Chain Token Control Attack}
  \label{fig:token}
\end{figure}

\subsection{Token Control Attacks}
Homogeneous DAOs face governance manipulation as centralized entities reserve voting power to sway decisions. In token-based systems like MakerDAO\cite{brennecke2022central} and Compound\cite{compound_governance}, whales dictate proposals, while liquidity incentives, as seen in Curve DAO, enable indirect control through token aggregation. In cross-chain DAOs, token control attacks are amplified by liquidity fragmentation and governance token disparities across chains. Attackers can exploit multi-chain arbitrage, acquiring undervalued tokens to influence decisions on chains with higher stakes. Manipulated cross-chain oracles can distort voting power, enabling attackers to gain control with minimal investment. The complexity of tracking token flows across chains makes detection and mitigation more challenging than in single-chain environments.

An attack scenario (Figure \ref{fig:token}) typically unfolds when an attacker accumulates governance tokens from multiple chains, exploiting price differentials and weaker security policies. They then use these tokens to either pass a malicious governance proposal or block an essential upgrade. Given the asynchronous nature of cross-chain decision-making, the attacker can execute token swaps or dumps before their actions trigger alarms, leaving the DAO with an imbalanced governance structure that may take weeks or months to rectify. This attack ultimately erodes trust in the governance system, discouraging participation from legitimate stakeholders.  

\subsection{HCI Attacks}
HCI attacks in DAOs exploit poor UI/UX, enabling governance manipulation through deceptive interfaces, misleading votes, or phishing. Attackers exploit complex processes to deceive users into approving malicious proposals or signing transactions that delegate voting power. Cross-chain DAOs increase governance complexity by decentralizing decision-making across multiple interfaces, raising the risk of errors or manipulation. Malicious actors can exploit cross-chain governance interfaces to misrepresent vote outcomes or hijack transactions via subtle UI changes. Additionally, attackers can inject delays or misleading confirmation messages through inter-chain messaging protocols, enabling social engineering attacks that compromise key governance decisions. 

\begin{figure}[htbp!]
  \centering
  \includegraphics[width=0.9\columnwidth]{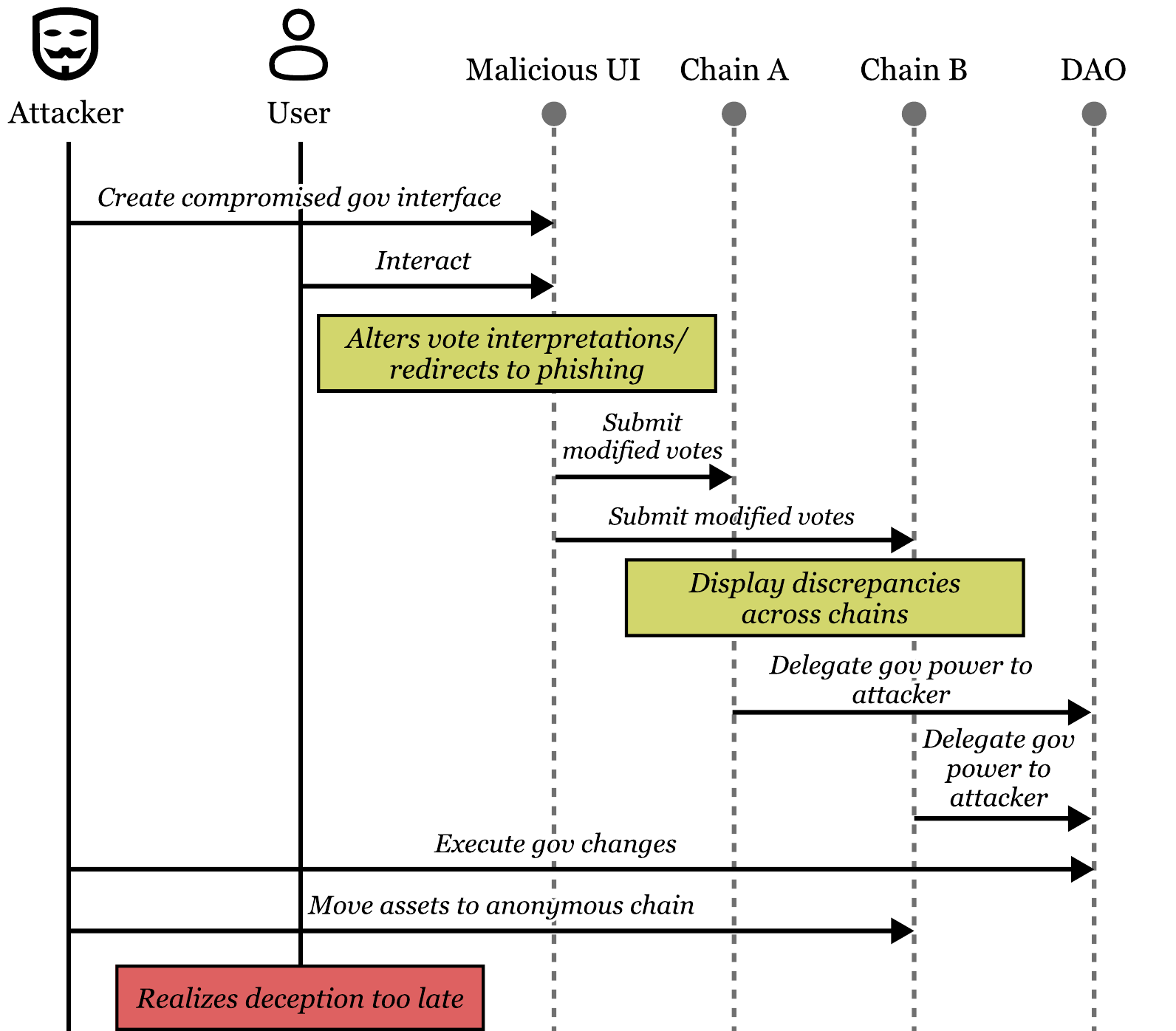}
  \caption{Cross-Chain HCI Attack}
  \label{fig:hci}
\end{figure}

An HCI attack begins with an attacker crafting a malicious governance interface that subtly alters vote interpretations or redirects users to a phishing portal (Figure \ref{fig:hci}). Unsuspecting DAO participants interact with the compromised interface, believing they are voting on a legitimate proposal, while in reality, their actions delegate governance power to the attacker. In a cross-chain scenario, the attack is extended by exploiting discrepancies in how different chains display governance data, creating confusion and misrepresentation. By the time users realize they have been deceived, governance changes may have already been executed, and the attacker has likely moved assets to an anonymous chain to evade detection.

\begin{figure*}[t]
  \centering
  \includegraphics[width=0.88\textwidth]{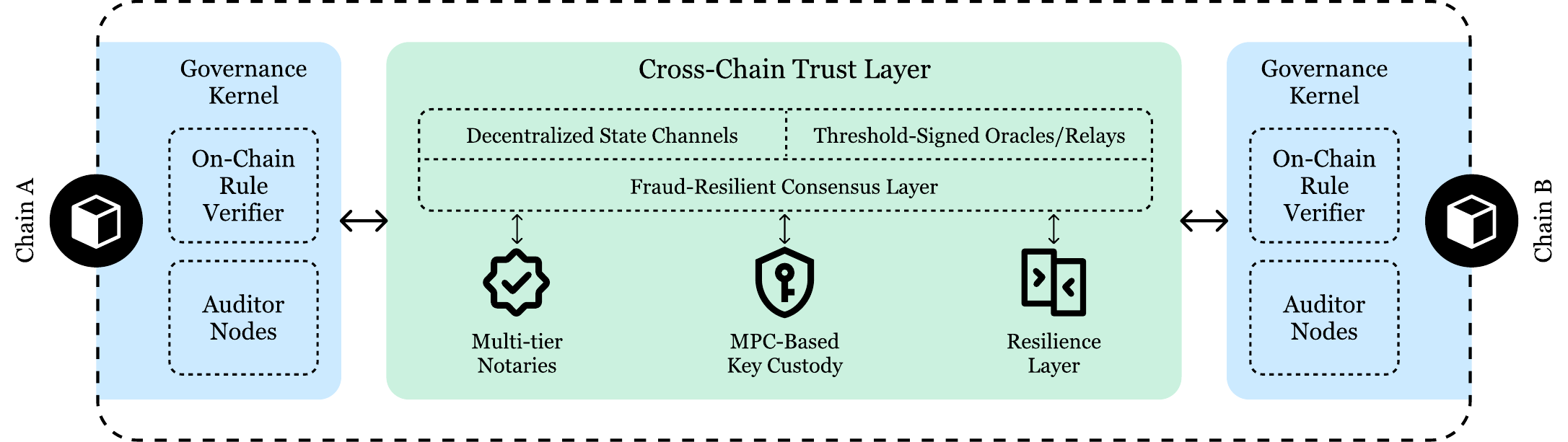}
  \caption{Overview of the Proposed Cross-Chain DAO Governance Framework}
  \label{fig:architecture}
\end{figure*}

\subsection{Code and Protocol Vulnerabilities}
Smart contract vulnerabilities in single-chain DAOs, such as reentrancy, misconfiguration, and Oracle manipulation, facilitate attacks. The 2016 DAO hack exposed reentrancy risks\cite{dhillon2017dao}, whereas Compound's exploits revealed governance flaws \cite{defiant_easyfi_hack}. Even audited contracts are vulnerable to unforeseen attacks.  

In a cross-chain environment, vulnerabilities arise from the complexity of interoperability protocols and reliance on external bridges and oracles. Attackers exploit inconsistencies in smart contract execution across chains, taking advantage of differences in execution order, time delays, or message propagation. Cross-chain bridges, like those exploited in the Wormhole\cite{behnke2022wormhole} and Ronin hacks\cite{behnke2024ronin}, are particularly vulnerable, introducing additional attack surfaces. Dependency on oracles also exposes DAOs to manipulated governance if attackers inject false data, causing erroneous decisions across chains.

Code-level attacks on cross-chain DAOs often involve exploiting governance rule inconsistencies between chains. An attacker may exploit a weaker chain with lax security policies to execute a governance action, then relay it to the primary chain through the interoperability protocol. By the time the primary chain processes the update, assets may be withdrawn or harmful changes made. Additionally, bridge manipulation can fragment consensus, preventing timely governance rollbacks and leaving the DAO compromised.

\section{Cross-Chain DAO Governance Framework}
\label{sec:framework}
We propose a modular, multi-layered theoretical framework for cross-chain DAO governance to ensure secure, verifiable, and attack-resilient governance across blockchains. The high-level overview of the framework is shown in Figure \ref{fig:architecture}.

\subsection{Governance Kernel and On-Chain Rule Verification}
The core of our proposed governance framework is the \textbf{Governance Kernel}, which would operate independently on each blockchain. This module would enforce governance rules through deterministic state machines, process voting mechanisms, and interact with smart contracts to execute DAO decisions. For compliance with predefined governance constraints, we propose an \textbf{On-Chain Rule Verifier} that would validate all governance actions against formal verification models before execution. Additionally, \textbf{Auditor Nodes} would function as decentralized monitors in a Byzantine fault-tolerant configuration, continuously analyzing governance activities for anomalies, potential attacks, or manipulations.

\subsection{Cross-Chain Trust Layer}
To securely extend governance across multiple blockchains, we propose a \textbf{Cross-Chain Trust Layer} that would ensure consistency and authenticity in cross-chain decision-making. This module would comprise multiple sub-components designed to handle secure governance interactions between chains.

\subsubsection{Decentralized State Channels}
State channels would provide an off-chain mechanism for executing governance decisions efficiently. By utilizing state channels with optimistic execution models, DAOs could achieve \textit{fast finality} in governance transactions without incurring the delays and costs of on-chain execution. Once consensus is reached, the governance actions would be committed on-chain in a cryptographically verifiable manner.

\subsubsection{Threshold-Signed Oracles and Relays}
To ensure secure transmission of governance data across chains, we propose \textbf{Threshold-Signed Oracles and Relays}. These oracles would use \textit{threshold cryptography (t,n)}\cite{de1994share}, requiring a quorum of trusted relays to sign off on governance updates before they are accepted. This design would prevent centralization risks and mitigate the possibility of oracle corruption or governance tampering.

\subsection{Fraud-Resilient Consensus Layer}
A key aspect of our governance security proposal is the \textbf{Fraud-Resilient Consensus Layer}, which would protect against bribery, governance takeovers, and collusion attempts. This layer would integrate \textit{Byzantine Fault-Tolerant (BFT) mechanisms} \cite{bashir2022blockchain} with stake-weighted validation and fraud detection models to penalize malicious actors attempting to compromise governance.

\subsection{Multi-Tier Trust Anchors and MPC-Based Key Custody}
To eliminate single point of failure in governance execution, we propose a \textbf{Multi-Tier Trust Anchor System}. This system would consist of decentralized \textbf{Notaries} to verify governance transactions before execution. Furthermore, we introduce \textbf{MPC-Based Key Custody}, which would ensure that governance-related cryptographic keys are managed through \textit{Multi-Party Computation}. This would prevent unauthorized access, unilateral governance takeovers, and key theft by distributing control across multiple independent parties.

\subsection{Resilience Layer and Dispute Resolution}
To address governance failures, attacks, and disputes, we propose a \textbf{Resilience Layer}. This module would provide mechanisms for \textit{time-locked decision reversals, emergency governance overrides, and multi-tiered dispute resolution}. If a governance attack is detected, validators and community stakeholders could initiate rollbacks before irreversible damage occurs. The system would also support \textbf{progressive arbitration}, starting with validator-led mediation and escalating to decentralized community voting when necessary.

\section{Conclusion and Future Directions}
\label{sec:conc}
In this paper, we analyzed cross-chain DAO governance attack vectors and proposed a theoretical framework to address these challenges. By categorizing attack vectors such as bribery, token control exploits, HCI, and code vulnerabilities, this paper laid the groundwork for understanding the threats to cross-chain DAOs. Our framework integrated threshold cryptography, MPC-based key custody, and time-locked decision reversals to defend against malicious manipulation while preserving the decentralized nature of DAO governance.

While our framework provides a strong foundation for secure cross-chain governance, several challenges remain. We will empirically evaluate the framework through prototype implementations to assess performance, latency, and deployment. We will also develop formal security proofs under various threat models to bolster confidence in its resilience. Additionally, we will explore dynamic trust adjustments, privacy-preserving technologies like zero-knowledge proofs, and economic incentive mechanisms to further enhance the framework. As regulatory requirements evolve, we will work on developing compliance-aware governance mechanisms. By addressing these challenges, this framework will help ensure that cross-chain DAOs remain secure, transparent, and resilient.

\bibliographystyle{IEEEtran}
\bibliography{IEEEabrv,references}

\end{document}